%modified due to referee Sept 24
%added references Sept 6
%Noureddine correction Sept 19
\input harvmac
\input epsf
\def\frak#1#2{{\textstyle{{#1}\over{#2}}}}
\def\frakk#1#2{{{#1}\over{#2}}}

\def\pa{\partial}
\def\semi{;\hfil\break}

\def\ptil{\tilde\phi}

\def\DRED{\ifmmode{{\rm DRED}} \else{{DRED}} \fi}
\def\DREDp{\ifmmode{{\rm DRED}'} \else{${\rm DRED}'$} \fi}
\def\NSVZ{\ifmmode{{\rm NSVZ}} \else{{NSVZ}} \fi}

\def\thbar{{\overline\theta}{}}

\def\Dbar{{\overline D}{}}

\def\Tr{\hbox{Tr}}

\def\hbar{{\overline h}{}}
\def \newin{\leftskip = 40 pt\rightskip = 40pt}

\def \out{\leftskip = 0 pt\rightskip = 0pt}
{\nopagenumbers
\line{\hfil LTH 518} 
\line{\hfil hep-th/0109015}
\line{\hfil Revised Version}
\vskip .5in
\centerline{\titlefont Ultraviolet properties of noncommutative }
\centerline{\titlefont non-linear 
$\sigma$-models in two dimensions}
\vskip 1in
\centerline{\bf I.~Jack${}^*$, D.R.T.~Jones${}^*$ and 
N.~Mohammedi${}^{\dagger}$}
\medskip
\centerline{\it ${}^*$Dept. of Mathematical Sciences,
University of Liverpool, Liverpool L69 3BX, UK}
\vskip .3in
\centerline{\it ${}^{\dagger}$Laboratoire de Math\'ematiques et
Physique Th\'eorique, Universit\'e Fran\c cois Rabelais, }
\centerline{\it F-37200 Tours, France}
\vskip .3in
We discuss the ultra-violet properties of bosonic and supersymmetric 
noncommutative non-linear $\sigma$-models in two dimensions, both with 
and without a Wess-Zumino-Witten term. 
  
\Date{September 2001}

There has been a great deal of recent interest in  noncommutative (NC)
quantum field theories, stimulated by their connection with string theory
and   $M$-theory; for a review and comprehensive list of references see
Ref.~\ref\DouglasBA{ M.R.~Douglas and N.A.~Nekrasov, hep-th/0106048}. Most
of this interest has focussed on four-dimensional theories. However, 
since two-dimensional theories have often been used as laboratories for
investigating general properties of quantum field theories, it is natural 
to extend the discussion to this arena. Two-dimensional non-commutative
non-linear $\sigma$-models have been discussed in
Refs.~\ref\DabrowskiMY{L.~Dabrowski, T.~Krajewski and G.~Landi, Int.\ J.\
Mod.\ Phys.\ B{\bf 14} (2000) 2367}\ref\GirottiGS{
H.O.~Girotti, M.~Gomes, V.O.~Rivelles and A.J.~da Silva,
hep-th/0102101}. A particularly interesting case to consider,
by virtue of its conformal invariance properties, is the Wess-Zumino-Witten
(WZW) model. This has been studied in the NC case in 
Refs.~\DabrowskiMY\ref\MorenoXU{ E.F.~Moreno and F.A.~Schaposnik, 
JHEP {\bf 0003} (2000) 032;
Nucl.\ Phys.\ B{\bf 596} (2001) 439\semi
C.~Nunez, K.~Olsen and R.~Schiappa,
JHEP {\bf 0007} (2000) 030}. The NC WZW term is 
also discussed in Ref.~\ref\ChuBZ{C.~Chu,
Nucl.\ Phys. B{\bf 580} (2000) 352}\ and 
the Kac-Moody algebra associated with the NC WZW model has been investigated in 
Ref.~\ref\GhezelbashPZ{A.M.~Ghezelbash and S.~Parvizi, Nucl.\ Phys.\
B{\bf 592} (2001) 408}. Moreover, its 
renormalisation has been carried out at one-loop order\ref\FurutaEP{ K.~Furuta
and T.~Inami, Mod.\ Phys.\ Lett.\ A{\bf 15} (2000) 997}. 
Our purpose in this paper is to continue the program of
perturbative investigation of the NC WZW model, and also the NC version
of the principal chiral model (i.e. the theory defined on a group manifold 
without the WZW term). We show how results for the NC $U_N$ WZW, and also
principal chiral, model may be 
obtained from the leading-$N$ term in the corresponding result for the 
commutative $SU_N$ theory.

Firstly we discuss the elements of the construction of NC field theories.
The algebra of functions  on a noncommutative space 
is isomorphic to the algebra of functions on a commutative 
space with coordinates $x^{\mu}$, with the product $f*g (x)$ 
defined as follows
\eqn\prodnc{
f*g (x) = e^{-i\Theta^{\mu\nu}
\frakk{\pa}{\pa\xi^{\mu}}\frakk{\pa}{\pa\eta^{\nu}}}
f(x+\xi)g(x+\eta)|_{\xi,\eta\to 0},}
where $\Theta$ is a real antisymmetric matrix.
Quantum field theories analogous to the corresponding commuting 
theories are  now straightforward to define, with $*$-products replacing 
ordinary products. In particular the 
noncommutative two-dimensional Wess-Zumino-Witten (WZW)
model is defined by
\eqn\WZWdef{
S=-{1\over{4\lambda^2}}\int_{\Sigma}d^2x\Tr\left(\pa_{\mu}gg^{-1}\pa^{\mu}g
g^{-1}\right)_*+{k\over{24\pi}}\int_Bd^3x\epsilon^{\mu\nu\rho}
\Tr\left(g^{-1}\pa_{\mu}gg^{-1}\pa_{\nu}gg^{-1}\pa_{\rho}g\right)_*,}
where as usual $\Sigma$ is the boundary of a three-dimensional manifold $B$,
and $g$ is a map from $\Sigma$ (or its extension $B$) into $U_N$.
(Note that $SU_N$ is not a group under the $*$-product, whereas $U_N$ is.)
$\epsilon^{\mu\nu\rho}$ is the three-dimensional alternating symbol.
A subscript $*$ in Eq.~\WZWdef\ indicates that every product of fields 
within the corresponding brackets is
a $*$-product. We assume that the co-ordinates $x^0$, $x^1$ on the 
worldsheet are non-commutative, but the extended co-ordinate $x^2$ on the 
manifold $B$ commutes with the others.
The group-valued field $g$ is defined as
\eqn\goupdef{
g=\exp_*(i\phi)=1+i\phi-{1\over{2!}}\phi*\phi+\ldots,}
where $\phi$ is in the Lie algebra of $U_N$. $\phi$ can be expanded as
\eqn\fielddef{
(\phi)^A_B=\phi_a(T_a)^A_B, \quad a=0,1,\ldots N^2-1, \quad A,B=1,\ldots N}
where $T_a$, $a=1,\ldots N^2-1$ are the generators of $SU_N$, 
$T_0=\sqrt{2\over{N}}1_N$, and with our conventions
\eqn\conv{
\Tr(T_aT_b)=2\delta_{ab}, \quad[T_a,T_b]=2if_{abc}T_c.}
The $U_N$ structure constants $f_{abc}$ are totally antisymmetric,
with $f_{0ab}=0$ and 
$f_{abc}$, $a=1,\ldots N^2-1$ being the structure constants of $SU_N$.
The commutative version of the theory is the sum of the commutative
$SU_N$ theory together with a free scalar field. Later on we compare the 
$\beta$-function for $\lambda$ in Eq.~\WZWdef\ with the corresponding 
$\beta$-function for the commutative $SU_N$ theory.

The ultra-violet properties of the NC WZW model may be investigated using the
background field method. We expand the field $g$ around a classical background 
$g_c$ as $g=g_c*g_q$, and express $g_q$ in terms of a quantum fluctuation 
$\pi$ as
\eqn\quandef{
g_q=\exp_*(i\lambda\pi).}
The expansion of the action may then be effected 
straightforwardly\ref\bos{M.~Bos, Ann. Phys. {\bf181} (1988) 177}; we
readily obtain
\eqn\PW{\eqalign{
S(g)=&S(g_c)\cr
&+{1\over{2\lambda^2}}\int_{\Sigma}d^2xP^{\mu\nu}
\Tr\left[e^{i\lambda\pi}\pa_{\mu}e^{-i\lambda\pi}g_c^{-1}\pa_{\nu}g_c
-i\lambda\pa_{\mu}\pi\int_0^1dte^{-it\lambda\pi}\pa_{\nu}e^{it\lambda\pi}
\right]_*,\cr}}
where
\eqn\projd{
P^{\mu\nu}=\eta^{\mu\nu}-{k\lambda^2\over{4\pi}}\epsilon^{\mu\nu},}
with $\epsilon^{\mu\nu}$ the two-dimensional alternating symbol,
and then derive an expansion in terms of $\pi$ by using
\eqn\gqexp{
\exp_*(i\lambda\pi)_*\pa_{\mu}[\exp_*(-i\lambda\pi)]=-i\lambda\pa_{\mu}\pi+
\frakk{(-i\lambda)^2}{2!}[\pa_{\mu}\pi,\pi]_*
+\frakk{(-i\lambda)^3}{3!}[[\pa_{\mu}\pi,\pi],\pi]_*+\ldots}
(together with a similar relation with $\lambda\rightarrow-t\lambda$).
The difference between noncommutative and commutative theories at the level of
Feynman diagrams is that in the NC case, Feynman diagrams can acquire 
momentum dependent phase factors arising from the $*$-product. If 
such a factor contains a loop momentum, the UV divergence for that loop is
suppressed. 
Since the $\pi$ are adjoint fields in $U_N$, the detailed discussion is
simplified by using the diagrammatic 
notation 
originally introduced by 't Hooft
\ref\tHooftJZ{
G.~'t Hooft,
%``A Planar Diagram Theory For Strong Interactions,''
Nucl.\ Phys.\ B{\bf 72} (1974) 461\semi
%%CITATION = NUPHA,B72,461;%%
P.~Cvitanovic, P.G.~Lauwers and P.N.~Scharbach,
%``The Planar Sector Of Field Theories,''
Nucl.\ Phys.\ B{\bf 203} (1982) 385
%%CITATION = NUPHA,B203,385;%%
}, where we represent a $\pi^A_B$ propagator
by a double line as in Fig.~1, the arrow pointing towards the upper index. 
\vskip 20pt
\epsfysize= 0.25in
\centerline{\epsfbox{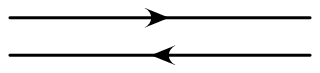}}
\vskip 10pt\newin
{\it \noindent Fig.~1:
The propagator for an adjoint $U_N$ field}
\bigskip
\out

In terms of this notation, the phase factors cancel in planar graphs,
and hence they give exactly
the same contributions to the renormalisation-group (RG) functions 
($\beta$-functions and anomalous dimensions) in the noncommutative $U_N$
case as in the commutative $SU_N$ case. In non-planar graphs, however,
the phase factors do not cancel, so the corresponding Feynman integrals are
UV convergent (after subtraction of subdivergences) 
and they do not contribute.  This was
first  shown in the case of NC gauge  theories in Ref.~\ref\MinwallaPX{
S.~Minwalla, M.~Van Raamsdonk and N.~Seiberg, JHEP {\bf 0002}  (2000)
020}, but the same argument applies here. Now the
planar  contributions give the leading order in powers of $N$, and therefore
the NC $U_N$ result can be obtained from the commutative $SU_N$ version by
extracting the leading term in $N$. This simple connection between the
NC and commutative cases makes it straightforward to extend  results
from the commutative to the NC case. 

We start by considering conformal invariance properties of
the WZW model. At the critical point  
\eqn\crit{\lambda^2={4\pi\over{k}}} 
the NC WZW model becomes conformally
invariant, as discussed in Ref.~\GhezelbashPZ. In
the commutative case the result can be derived  straightforwardly
starting from  the commutative version of Eq.~\PW\bos (for the generalisation 
to an arbitrary parallelised manifold see 
Ref.~\ref\MukhiVZ{
S.~Mukhi,
%``Finiteness Of Nonlinear Sigma Models With Parallelizing Torsion,''
Phys.\ Lett.\ B {\bf 162} (1985) 345 
%%CITATION = PHLTA,B162,345;%%
}).  We sketch the
proof here.  Feynman diagrams are constructed with vertices
derived from the expansion of Eq.~\PW\ in terms of $\pi$; 
the propagator, derived from the
term in Eq.~\PW\ quadratic in $\pi$, is simply ${\eta^{\mu\nu}
\over{k^2}}$. Note that there are two sorts of vertex; those with one
derivative acting on a quantum field $\pi$ and one factor of 
$g_c^{-1}\pa_{\nu}g_c$, (Type A) and  those with two derivatives acting
on $\pi$ and no factors of  $g_c^{-1}\pa_{\nu}g_c$ (Type B). Each vertex
contains a factor of $P^{\mu\nu}$, although by symmetry only the 
$\eta^{\mu\nu}$ or the $\epsilon^{\mu\nu}$ in $P^{\mu\nu}$ 
contributes to the Type B vertices with even or
odd numbers of $\pi$s respectively. The contributions to the 
renormalisation of $\lambda$ arise from logarithmically divergent
diagrams, which contain two Type A vertices and an arbitrary number of
Type B vertices. (The WZW term is not renormalised
\ref\MukhiVY{
S.~Mukhi,
%``The Geometric Background Field Method, 
%Renormalization And The Wess-Zumino Term In Nonlinear Sigma Models,''
Nucl.\ Phys.\ B {\bf 264} (1986) 640 
%%CITATION = NUPHA,B264,640;%%
}.)
If one is using dimensional regularisation, it is
necessary to have a  prescription for products of $\epsilon$ tensors,
valid in $d\ne2$ dimensions. The simplest is to
define\ref\HTJ{C.M.~Hull and P.K.~Townsend, Phys.Lett. B{\bf 191} (1987)
115\semi R.R.~Metsaev and A.A.~Tseytlin, Phys. Lett. B{\bf 191} (1987)
354\semi H.~Osborn, Ann. Phys. {\bf200} (1990) 1}
\eqn\epsdf{\epsilon^{\mu\rho}\epsilon_{\rho\nu}= \delta^{\mu}{}_{\nu},}}
the contraction here being effected by the  $d$-dimensional metric.
(This definition leads to conformal invariance at the  critical point
(Eq.~\crit) without the necessity of additional finite counter-terms.) 
Crucial is that we now have in $d$ dimensions 
\eqn\proj{
P^{\mu\rho}P^{\nu}{}_{\rho}=\left[1-\left({k\lambda^2\over{4\pi}}\right)^2
\right]\eta^{\mu\nu}.} After performing all the Feynman integrals and
implementing all the resulting tensor algebra, the final result is
proportional to 
$P^{\mu\rho}P^{\nu}{}_{\rho}\Tr[\pa_{\mu}g_cg_c^{-1}\pa_{\nu}g_cg_c^{-1}]_*$.
At the critical point Eq.~\crit\ we have
$P^{\mu\rho}P^{\nu}{}_{\rho}=0$, and therefore there are no corrections
to $\lambda$. Clearly, since this proof  relies only on the tensor
structure of the vertices and not on the details of the Feynman
diagrams, it is unaffected by the additional phase factors present in
the NC case. Moreover, the proof in the commutative, and hence also the
NC case, is equally valid for the supersymmetric theory (which will be
defined explicitly later), as was emphasised in
Ref.~\ref\allen{R.W.~Allen, I.~Jack and D.R.T.~Jones, Z. Phys. C{\bf41} (1988)
323}.

Another result for the commutative bosonic $SU_N$ case, which can be proved
using conformal field theory arguments, is \ref\KnizhnikNR
\ref\KnizhnikNR{ V.G.~Knizhnik and A.B.~Zamolodchikov, Nucl.\ Phys.\
B{\bf 247} (1984) 83} 
\eqn\bderiv{ \left.{\pa\beta_{\lambda}
\over{\pa\lambda^2}}\right|_{\lambda^2={4\pi\over{k}}}
={4N\over{k+2N}}.} 
We can expand Eq.~\bderiv\ as  
\eqn\bexp{
\left.{\pa\beta_{\lambda}
\over{\pa\lambda^2}}\right|_{\lambda^2={4\pi\over{k}}} ={\lambda^2
N\over{\pi}}\left[1-{\lambda^2 N\over{2\pi}}+ \left({\lambda^2
N\over{2\pi}}\right)^2-\ldots\right].} 
This result can be interpreted as
a perturbative loop expansion. Each term is  leading order in $N$ for
the corresponding loop order. Therefore the result for  the NC $U_N$
theory will be identical, and we deduce that Eq.~\bderiv\ is also valid
for the $U_N$ NC WZW model. The $\beta$-function 
$\beta_{\lambda}$ for the commutative
WZW model has been computed up to three loops in Ref.~\ref\xi{Z-M. Xi, 
Phys.Lett.  B{\bf214} (1988) 204; Nucl. Phys. B{\bf314} (1989) 112}.
After specialising to  $SU_N$, the result is leading order in $N$ at
this order and hence the result is identical in the NC $U_N$ case. For
completeness, we quote it here: 
\eqn\betathree{
\beta_{\lambda}=-\lambda^2(1-\eta^2)\left[2\rho+2\rho^2(1-3\eta^2)+3\rho^3
(1-\frak{25}{3}\eta^2+10\eta^4) + \cdots\right],} 
where
$\eta={k\lambda^2\over{4\pi}}$ and $\rho={\lambda^2N\over{4\pi}}$. (In
fact three-loop results have also been given in  Ref.~\ref\KetovHW{
S.V.~Ketov, Theor.\ Math.\ Phys.\ {\bf 80} (1989) 709 [Teor.\ Mat.\
Fiz.\ {\bf 80} (1989) 56]\semi S.V.~Ketov, A.A.~Deriglazov and
Y.S.~Prager, Nucl.\ Phys.\ B{\bf 332} (1990) 447}, but apparently in
a different  renormalisation scheme.) It is easy to verify that 
Eq.~\betathree\ is compatible with Eqs.~\bderiv, \bexp; notice that 
in taking the derivative with respect of $\lambda^2$ of Eq.~\betathree, 
only the terms arising from hitting the $(1-\eta^2)$ factor survive 
because the result  is to be evaluated at $\eta^2 = 1$.  

Let us now turn to the supersymmetric 
case. The NC supersymmetric WZW model
has the superspace action
\ref\DiVecchiaEP{T.L.~Curtright and C.K.~Zachos,
%``Geometry, Topology And Supersymmetry In Nonlinear Models,''
Phys.\ Rev.\ Lett.\  {\bf 53} (1984) 1799 \semi
%%CITATION = PRLTA,53,1799;%%
S.J.~Gates, C.M.~Hull and M.~Rocek,
%``Twisted Multiplets And New Supersymmetric Nonlinear Sigma Models,''
Nucl.\ Phys.\ B {\bf 248} (1984) 157 \semi
%%CITATION = NUPHA,B248,157;%%
E.~Abdalla and M.C.~Abdalla,
Phys.\ Lett.\ B {\bf 152} (1985) 59\semi
P.~Di Vecchia et al,
Nucl.\ Phys.\ B{\bf 253} (1985) 701\semi
E.~Braaten, T.L.~Curtright and C.K.~Zachos,
%``Torsion And Geometrostasis In Nonlinear Sigma Models,''
Nucl.\ Phys.\ B {\bf 260} (1985)  630
%%CITATION = NUPHA,B260,630;%%
}
\eqn\WZWSUSY{
S_{\rm SUSY}={1\over{4\lambda^2}}\int d^2xd^2\theta\Tr\left[\Dbar
G^{-1}DG
\right]_*+{k\over{16\pi}}
\int d^3xd^2\theta \Tr\left[G^{-1}{dG\over{dt}}\Dbar G^+
\gamma_3DG\right]_*,}
where $t\equiv x^2$, and $\theta^{\alpha}$ are the Grassman co-ordinates 
and $G$ is now a superfield and a group element of $U_N$, 
defined in terms of a superfield $\Phi$ 
as $G=\exp_*(i\Phi)$. The supercovariant 
derivative $D$ is defined by
\eqn\superD{
D_{\alpha}={\pa\over{\pa\thbar^{\alpha}}}+i(\gamma^{\mu}\theta)_{\alpha}
\pa_{\mu},}
and $\gamma_3=\gamma^0\gamma^1$.
In the commutative supersymmetric $SU_N$ case $\beta_{\lambda}$ 
is given through three loops by
\eqn\betathreeSUSY{ \beta_{\lambda}=-2\rho\lambda^2(1-\eta^2),} 
i.e. the
two\ref\FridlingHC{
B.E.~Fridling and A.E.~van de Ven,
%``Renormalization Of Generalized Two-Dimensional Nonlinear Sigma Models,''
Nucl.\ Phys.\ B {\bf 268} (1986) 719\semi 
%%CITATION = NUPHA,B268,719;%%
D.R.T.~Jones, Phys. Lett. B{\bf192} (1987) 391}\
and three\ref\ket{S.V.~Ketov, Phys. Lett. B{\bf207} (1988) 140}   
loop contributions 
vanish\foot{In the torsion-free case this was shown for a general manifold 
in Refs.~\ref\AlvarezGaumeHN{
L.~Alvarez-Gaum\'e, D.Z.~Freedman and S.~Mukhi,
%``The Background Field Method And The 
%Ultraviolet Structure Of The Supersymmetric Nonlinear Sigma Model,''
Ann. Phys.\  {\bf 134} (1981) 85 
%%CITATION = APNYA,134,85;%%
}, \ref\AlvarezGaumePS{
L.~Alvarez-Gaum\'e,
%``Three Loop Finiteness In Ricci Flat Supersymmetric Nonlinear Sigma Models,''
Nucl.\ Phys.\ B {\bf 184} (1981) 180 
%%CITATION = NUPHA,B184,180;%%
}}; this property clearly carries
over to the NC case. The corresponding result to Eq.~\bderiv\ in the 
supersymmetric case can be deduced from Ref.~\ref\fuchs{J.~Fuchs, 
Nucl. Phys. B{\bf286} (1987) 455}, namely
\eqn\bderivSUSY{\left.{\pa\beta_{\lambda}
\over{\pa\lambda^2}}\right|_{\lambda^2={4\pi\over{k}}}
={4N\over{k}},}
in other words the result for $\left.{\pa\beta_{\lambda}
\over{\pa\lambda^2}}\right|_{\lambda^2={4\pi\over{k}}}$ is one-loop exact.
This result will clearly be equally valid in the NC case.
Eq.~\bderivSUSY\ is consistent with the perturbative results through three 
loops, and 
predicts that $\beta_{\lambda}$ at four loops and beyond should be proportional 
to $(1-\eta^2)^2$. 
Results have been presented  at the 4-loop level\ref\DeriglazovTM
\ref\DeriglazovTM{ A.A.~Deriglazov and S.V.~Ketov, Nucl.\ Phys.\ B{\bf
359} (1991) 498} for a general $N=1$ supersymmetric $\sigma$-model with
torsion, but the specialisation to the group manifold case appears to be 
incorrect and we have been unable to verify Eq.~\bderivSUSY\ at this level. 

In view of this uncertainty at four loops in the case of the WZW model, 
we now turn to the case of the NC 
$\sigma$-model defined on a group manifold without a WZW  term--i.e. the
NC version of the principal chiral model. Once again, we can  obtain the
NC results for the $\beta$-functions for the group $U_N$ simply by  picking
out the leading $N$ behaviour of the corresponding commutative results
for $SU_N$. 
We start with the bosonic case. Results are available for the general
bosonic  $\sigma$ model at two
\ref\FriedanJF{
D.~Friedan,
%``Nonlinear Models In Two Epsilon Dimensions,''
Phys.\ Rev.\ Lett.\  {\bf 45} (1980) 1057 
%%CITATION = PRLTA,45,1057;%%
}\AlvarezGaumeHN, three\ref\foakes{S.J.~Graham, Phys.
Lett. B{\bf197} (1987) 543\semi A.P.~Foakes and N.~Mohammedi, Phys.
Lett. B{\bf 198} (1987) 359; Nucl. Phys. B{\bf306} (1988) 343} and 
four\ref\JackVP{ I.~Jack, D.R.T.~Jones and N.~Mohammedi, Nucl.\ Phys.\
B{\bf 322} (1989) 431}\ 
loops,  expressed in terms of the Riemann tensor for
the target space metric. The results for the $SU_N$ case may be obtained
by substituting the appropriate Riemannn tensor; in terms of general 
co-ordinates $\ptil^k$ on the (commutative) group manifold, we have
\eqn\riedef{
R_{klmn}=e_k{}^ae_l{}^be_m{}^ce_n{}^df_{abe}f_{cde},} 
where $f_{abc}$
are the structure constants for $SU_N$ and $e_k{}^a$ are the  vielbeins
for the metric on the group manifold, defined by
\eqn\vieldef{e_k{}^ae_l{}^a=g_{kl}, \quad g^{kl}e_k{}^ae_l{}^b=\delta^{ab}.}
We find 
\eqn\bosres{
\beta_{\lambda}=-\lambda^2\left[2\rho+2\rho^2+3\rho^3+\rho^4\left(\frak{19}{3}
+\frak{12}{N^2}\zeta(3)\right)+\ldots\right].} 
We deduce that the result
in the NC $U_N$ case is given by 
\eqn\bosNCres{
\beta_{\lambda}=-\lambda^2\left[2\rho+2\rho^2+3\rho^3+\frak{19}{3}\rho^4
+\ldots\right].}

Finally we turn to the case of the supersymmetric principal chiral model. 
As we already know from our earlier discussion of the WZW model,
in this case the first non-zero contribution
to the $\beta$-function beyond one loop appears at four loops
\ref\GrisaruPX{
M.T.~Grisaru, A.E.~van de Ven and D.~Zanon,
Phys.\ Lett.\ B{\bf 173} (1986) 423;
Nucl.\ Phys.\ B{\bf 277} (1986) 388;
Nucl.\ Phys.\ B{\bf 277} (1986) 409}.
The result in the commutative $SU_N$ case may be obtained by substituting 
Eq.~\riedef\ into
the general results given in Ref.~\GrisaruPX, or, more easily, by 
recalling\JackVP\ 
that the four-loop $N=1$ supersymmetric result is identical to the part of the 
four-loop bosonic result involving $\zeta(3)$. We then see from Eq.~\bosres\
that there is no leading contribution at four loops in the supersymmetric case.
We deduce that $\beta_{\lambda}$ for the NC $N=1$ supersymmetric $U_N$  
principal chiral model vanishes from two through four loops.

In conclusion: we have established by perturbative arguments that the NC WZW 
$U_N$ model (bosonic or supersymmetric) is all-orders finite at the critical 
point. We have 
pointed out that results for the NC $U_N$ WZW or principal chiral model
can be derived from the corresponding commutative $SU_N$ result by 
extracting the leading-$N$ term. 
This immediately led to Eq.~\bderiv\ for the bosonic NC WZW $U_N$ model 
and Eq.~\bderivSUSY\ for the supersymmetric NC WZW $U_N$ model,
together with the 
three-loop results Eq.~\betathree\ for the bosonic theory 
and Eq.~\betathreeSUSY\ for the corresponding supersymmetric theory.
In the case of the bosonic NC $U_N$ principal chiral model we have given the 
$\beta$-function up to four loops in Eq.~\bosNCres; and we have deduced that
in the supersymmetric version of this theory,
the $\beta$-function vanishes from two through four 
loops. This tempts us to speculate that $\beta_{\lambda}$ may be one loop exact 
in this case, at least when using standard dimensional reduction.
It is not clear however how to determine whether 
there are any general reasons why this result should persist at higher orders.
The generally covariant methods used in the calculation of $\beta$-functions for
general $\sigma$ models as in Ref.~\GrisaruPX\ are not very well adapted for 
extracting the leading $N$ behaviour; on the other hand,
we have repeated the 4-loop $N=1$ 
computation using the non-covariant expansion as in Eq.~\PW, and extracted 
the contributions which are planar in terms of  
't Hooft's double line notation, but this does not seem to afford any general
insights.

\bigskip\centerline{{\bf Acknowledgements}}

IJ thanks the Laboratoire de Math\'ematiques et
Physique Th\'eorique, Universit\'e Fran\c cois Rabelais, Tours
for hospitality while part of this work was carried out. DRTJ was
supported in part by a PPARC Senior Fellowship.

\listrefs
\bye